\documentclass[prb,letterpaper,aps,floatfix,twocolumn]{revtex4-1}
\usepackage{graphicx}
\usepackage{amsmath}
\usepackage{subfigure}
\newcommand{\beq}{\begin{equation}}
\newcommand{\eeq}{\end{equation}}
\newcommand{\bk}{{{\bf{k}}}}

\newcommand{\br}{{{\bf{r}}}}

\newcommand{\bA}{{\bf{A}}}
\newcommand{\bB}{{\bf{B}}}
\newcommand{\bE}{{\bf{E}}}

\newcommand{\bq}{{\bf{q}}}

\newcommand{\beqa}{\begin{eqnarray}}
\newcommand{\eeqa}{\end{eqnarray}}
\newcommand{\pdg}{{\vphantom \dag}}
\newcommand{\dg}{{\dag}}
\newcommand{\bnabla}{{\boldsymbol \nabla}} 
\newcommand{\bsigma}{{\boldsymbol \sigma}}
\newcommand{\upa}{\uparrow}
\newcommand{\da}{\downarrow} 

\begin{document}
\title{Topological response in ferromagnets}
\author{A.A. Burkov}
\affiliation{Department of Physics and Astronomy, University of Waterloo, Waterloo, Ontario 
N2L 3G1, Canada} 
\date{\today}
\begin{abstract}
We present a theory of the intrinsic anomalous Hall effect in a model of a doped Weyl semimetal, 
which serves here as the simplest toy model of a generic three-dimensional metallic ferromagnet
with Weyl nodes in the electronic structure. 
We analytically evaluate the anomalous Hall conductivity as a function of doping, which allows us to explicitly 
separate the Fermi surface and non Fermi surface contributions to the Hall conductivity by carefully evaluating the 
zero frequency and zero wavevector limits of the corresponding response function. We show that this separation 
agrees with the one suggested a long time ago in the context of the quantum Hall effect by Streda.
\end{abstract}
\maketitle
\section{Introduction}
\label{sec:1}
There is a significant current interest in topologically-nontrivial properties of the electronic structure of solid materials. 
While the recent explosive expansion of this field is tied to the discovery of topological insulators (TI),~\cite{TI}
there is a growing realization that topologically-nontrivial properties occur in a significantly broader class of systems, not 
exhausted by TI. 
In particular, as has been anticipated some time ago,~\cite{Volovik} and fully realized recently, certain gapless systems, 
exhibiting point touching nodes between nondegenerate bands, or Weyl nodes, are topologically nontrivial.~\cite{Wan11,Ran11,Burkov11,Xu11}
Such {\em Weyl semimetals} realize massless chiral fermions in a condensed matter setting, and possess unique physical properties. 
Most of these may be thought of as being consequences of chiral anomaly, a phenomenon that is characteristic of chiral fermions in 
odd spatial dimensions (and thus not observed in graphene, for example).~\cite{Aji11,Burkov12,Grushin12,Goswami13,Pesin13}
The very recent experimental realization of {\em Dirac semimetals}~\cite{Cava13,Shen13,Hasan13} paves the way for the realization of 
Weyl semimetals in the near future.

Some of the interesting physics of Weyl semimetals, however, in fact occurs in an even broader class of materials. 
In particular, the electronic structure of (almost) any three-dimensional (3D) metallic ferromagnet possesses Weyl nodes, even if they 
are generally not aligned with the Fermi level, as would be the case in a Weyl semimetal.~\cite{Burkov13} 
These nodes play an important role in the phenomenon of the anomalous Hall effect (AHE),~\cite{Karplus54,Niu99,MacDonald02,Nagaosa02,Fang03,MacDonald04,Haldane04,Ong04,Nagaosa10}
the interest in which has grown sharply in the last decade with the realization~\cite{Niu99,MacDonald02} that topological properties of the electronic structure play an important, if not defining, 
role in it, making it a close relative of the quantum Hall effect. 

In this paper, we study a simple model of a Weyl semimetal, introduced by us before,~\cite{Burkov11} focusing on the dependence of the intrinsic AHE 
on doping. This model of a doped Weyl semimetal will serve here as the simplest possible toy model of the electronic structure of a 3D metallic ferromagnet with Weyl nodes, 
in which nearly everything can be calculated analytically, thus making our arguments especially clear.
In particular, we will demonstrate explicitly that it is possible to separate contributions to the intrinsic anomalous Hall conductivity into Fermi surface and non Fermi surface 
contributions by taking the low-frequency and long-wavelength limits of the corresponding response function in different order, which corresponds to evaluating 
either a transport or a thermodynamic equilibrium property. 
We demonstrate that Weyl nodes significantly affect the non Fermi surface contribution to the AHE.   
\section{Theta term in doped Weyl semimetal}
\label{sec:2}
We will consider a specific realization of a doped Weyl semimetal, based on TI-NI  (NI stands for normal insulator) heterostructure, doped with magnetic 
impurities.~\cite{Burkov11} The main advantage of this model of Weyl semimetal is its simplicity: all calculations can be done analytically in this case. 
Our final results, however, will be of general relevance and to some extent independent of the specifics of a particular model of a metallic ferromagnet. 
We start from the imaginary time action of electrons in the Weyl semimetal, coupled to electromagnetic field
\beq
\label{eq:1}
S = \int d \tau d^3 r \left\{ \Psi^\dg(\br, \tau) \left[\partial_{\tau} - \mu + i e A_0 (\br, \tau) + \hat H\right] \Psi^\pdg(\br, \tau)\right\}, 
\eeq
where $A_0(\br, \tau)$ is the scalar potential and 
\beq
\label{eq:2}
\hat H = v_F \tau^z (\hat z \times \bsigma) \cdot \left(- i \bnabla + e \bA \right) + \hat \Delta + b \sigma^z,
\eeq
is the Hamiltonian of noninteracting electrons in Weyl semimetal, minimally coupled to the vector potential $\bA$. 
We will ignore the $z$-component of the vector potential as it will not play any role in what follows. 
Throughout this paper we will use the units in which $\hbar = c =1$. 
$\hat \Delta = \Delta_S \tau^x \delta_{i,j} + \Delta_D (\tau^+ \delta_{j,i+1} + h.c.)/2$ is the operator, describing tunneling of electrons 
between the top (pseudospin $\upa$) and bottom (pseudospin $\da$) surfaces of same or neighboring TI layers in the 
heterostructure.~\cite{Burkov11} We will assume that $\Delta_S$ is positive for concreteness, while $\Delta_D$ can have any sign. 
The $b \sigma^z$ term describes the spin splitting due to uniform magnetization in the $z$-direction. 

Turning the electromagnetic field off for a moment, the Hamiltonian $\hat H$ is easily diagonalized in several simple steps. 
After Fourier transforming to momentum space and a canonical transformation of the spin and pseudospin variables $\sigma^{\pm} \rightarrow \tau^z \sigma^{\pm}$, 
$\tau^{\pm} \rightarrow \sigma^z \tau^{\pm}$, we obtain
\beq
\label{eq:3}
H(\bk) = v_F (\hat z \times \bsigma) \cdot \bk + [b + \hat \Delta(k_z)] \sigma^z. 
\eeq
The tunneling operator $\hat \Delta(k_z)$ can now be diagonalized separately. Its eigenvalues are given by $t \Delta(k_z)$, where $t = \pm$ and 
$\Delta(k_z) = \sqrt{\Delta_S^2 + \Delta_D^2 + 2 \Delta_S \Delta_D \cos(k_z d)}$. 
The corresponding eigenvectors are
\beq
\label{eq:4}
|u^t_{k_z} \rangle = \frac{1}{\sqrt{2}}\left(1, t \frac{\Delta_S + \Delta_D e^{-i k_z d}}{\Delta(k_z)} \right). 
\eeq
Note that the eigenvectors of the tunneling operator depend only on $k_z$. 
Introducing $m_t(k_z) = b + t \Delta(k_z)$, we can now rewrite the momentum-space Hamiltonian Eq.~\eqref{eq:3} as
\beq
\label{eq:4.1}
H_t(\bk ) =  v_F (\hat z \times \bsigma) \cdot \bk + m_t(k_z) \sigma^z. 
\eeq

$H_t(\bk)$ may now be viewed in a more general way as the simplest description of the electronic structure of a 3D ferromagnet.
The ``mass term" $m_+(k_z)$ is always positive, while $m_-(k_z)$ may change sign from positive to negative at the Weyl nodes, 
whose locations along the $k_x = k_y = 0$ line are determined by the solutions of the equation
\beq
\label{eq:4.2}
b - \Delta(k_z) = 0.
\eeq
When $b = 0$, $m_{\pm}(k_z) = \pm \Delta(k_z)$ and Eq.~\eqref{eq:4.1} describes two pairs of Kramers-degenerate bands. 
Kramers degeneracy when $b = 0$ is the reason that a minimal model of a ferromagnet with spin-orbit interactions must include 
four bands. 

Diagonalizing $H_t(\bk)$ in the space of the spin operators, we obtain its eigenvalues
\beq
\label{eq:5}
\epsilon_{s t}(\bk) = s \sqrt{v_F^2 (k_x^2 + k_y^2) + m_t^2(k_z)} = s \epsilon_{t}(\bk),
\eeq
and eigenvectors
\beq
\label{eq:6}
|v^{st}_{\bk}\rangle = \frac{1}{\sqrt{2}} \left[ \sqrt{1 + s \frac{m_t(k_z)}{\epsilon_t(\bk)}}, - i s e^{i \varphi_\bk}  \sqrt{1 - s \frac{m_t(k_z)}{\epsilon_t(\bk)}} \right],
\eeq
where $s = \pm$ and $e^{i \varphi_\bk} = k_+/\sqrt{k_x^2 + k_y^2}$. 
The vortex-like structure of the spinor $|v^{s t}_{\bk} \rangle$ due to the presence of the $e^{i \varphi_\bk}$ phase factor is behind the ``topological" physical 
properties of this system that we will discuss below. 
The full eigenfunctions of $H$ thus have the form of a tensor product of two spinors
\beq
\label{eq:7}
|z^{s t}_{\bk} \rangle =  |v^{st}_\bk \rangle \otimes |u^t_{k_z} \rangle.
\eeq

We can now integrate out electron variables in Eq.~\eqref{eq:1} and obtain an effective action for the electromagnetic field, induced by coupling to the electrons. 
This action will contain two distinct kinds of contributions. The first kind will contain terms, proportional to $\bE^2$ and $\bB^2$, 
where $\bE$ and $\bB$ are the electric and magnetic fields. These terms describe the electric and magnetic polarizability of the material. 
The second kind contains the ``topological" contribution, which has 
the form of a ``3D Chern-Simons term" (which may, alternatively, be thought of as the $\bE \cdot \bB$ term, but with a spatially-dependent coefficient~\cite{Burkov12}). 
Adopting the Coulomb gauge $\bnabla \cdot \bA = 0$, the topological contribution to the 
electromagnetic field action takes the following form
\beq
\label{eq:8}
S = \sum_{\bq, i\Omega}\epsilon^{z 0 \alpha \beta} \Pi(\bq, i\Omega) A_0(-\bq, -i\Omega) \hat q_{\alpha} A_{\beta}(\bq, i \Omega),
\eeq 
where $\hat q_{\alpha} = q_{\alpha}/ q$ and summation over repeated indices is implicit. The $z$-direction in Eq.~\eqref{eq:8} is picked out by the 
magnetization $b$. 
The response function $\Pi(\bq, i\Omega)$ is given by
\beqa
\label{eq:9}
\Pi(\bq, i \Omega)&=& \frac{i e^2 v_F}{V}  \sum_{\bk} \frac{n_F[\xi_{s' t'}( \bk)] - n_F[\xi_{s t} (\bk + \bq)]}{i \Omega + \xi_{s' t'}(\bk) - \xi_{s t} (\bk + \bq)} \nonumber \\
&\times&\langle z^{s t}_{\bk + \bq} | z^{s' t'}_{\bk} \rangle \langle z^{s' t'}_{\bk}| \bsigma \cdot \hat q | z^{s t}_{\bk + \bq} \rangle,
\eeqa 
where $\xi_{s t}(\bk) = \epsilon_{s t}(\bk) - \epsilon_F$ and summation over repeated band indices $s, t$ is implicit.  

To evaluate $\Pi(\bq, i\Omega)$ explicitly it is convenient to rotate coordinate axes so that $\bq = q \hat x$ and assume that $\epsilon_F > 0$. The $\epsilon_F < 0$ result is 
evaluated analogously. 
There exist two kinds of contributions to $\Pi(\bq, i \Omega)$: {\em interband}, with $s \neq s'$, and {\em intraband}, with $s = s' = +$. 
Let us first evaluate the interband contributions. 
In this case we can set $i \Omega = 0$ in the denominator of Eq.~\eqref{eq:9} from the start.
Since the pseudospin part of the eigenvectors $|z^{s t}_{\bk} \rangle$ depends only on $k_z$, i.e. is independent of $\bq = q \hat x$, the scalar product of the 
pseudospin wavefunctions simply gives us $\delta_{t t'}$. 
We then note that $\langle z^{\pm t}_{\bk + \bq} | z^{\mp t}_{\bk} \rangle \rightarrow 0$ when $\bq \rightarrow 0$, while $\langle z^{\pm t}_{\bk + \bq} | \sigma^x | z^{\mp t}_{\bk} \rangle$ 
remains finite in this limit. To leading order in $\bq$ we can then expand $\langle z^{\pm t}_{\bk + \bq} | z^{\mp t}_{\bk} \rangle$ to first order in $\bq$ while setting $\bq = 0$ everywhere else. 
We obtain
\beq
\label{eq:10}
\langle z^{\pm t}_{\bk + \bq} | z^{\mp t}_{\bk} \rangle = \mp \frac{v_F q}{2 \epsilon_t(\bk) \sqrt{k_x^2 + k_y^2}} \left[ \pm i k_y + \frac{m_t(k_z)}{\epsilon_t(\bk)} k_x \right]. 
\eeq
and 
\beq
\label{eq:11}
\langle z^{\pm t}_{\bk} | \sigma^x | z^{\mp t}_{\bk} \rangle = \frac{1}{\sqrt{k_x^2 + k_y^2}} \left[\pm i k_x  - \frac{m_t(k_z)}{\epsilon_t(\bk)} k_y\right]. 
\eeq 
Substituting this into Eq.~\eqref{eq:9} and leaving only the terms that will survive angular integration, we obtain
\beq
\label{eq:12}
\Pi^{inter}(\bq, i\Omega) = \frac{e^2 v_F^2 q}{2 V} \sum_t  \sum_{\bk} \frac{1 - n_F[\epsilon_t(\bk) - \epsilon_F]}{\epsilon_t^3(\bk)} m_t(k_z). 
\eeq
Introducing $x = v_F^2 (k_x^2 + k_y^2)$, we can rewrite this as
\beqa
\label{eq:13}
&&\Pi^{inter}(\bq, i\Omega) = - \frac{e^2 q}{8 \pi^2} \sum_t \int_{-\pi/d}^{\pi/d} d k_z \int_0^{\infty} dx\,\, \Omega^t_z(x, k_z) \nonumber \\
&\times&[1 - n_F(\sqrt{x + m_t^2(k_z)} - \epsilon_F)], 
\eeqa
where 
\beq
\label{eq:14}
\Omega^t_z(x, k_z) = -\frac{m_t(k_z)}{2 [x + m^2_t(k_z)]^{3/2}}, 
\eeq
is the $z$-component of the Berry curvature of the $s = -, t$ bands. 
The first term in Eq.~\eqref{eq:13} is the contribution to $\Pi^{inter}(\bq, i\Omega)$ of the two $s = -$ bands, which are completely filled when $\epsilon_F > 0$. 
The second term is the contribution of the incompletely filled $s = +$ bands. 
Integrating over $x$, we obtain
\beqa
\label{eq:15}
&&\Pi^{inter}(\bq, i\Omega) = \frac{e^2 q}{8 \pi^2} \sum_t  \int_{-\pi/d}^{\pi/d} d k_z \,\, \textrm{sign} [m_t(k_z)]  \nonumber \\
&-&\frac{e^2 q}{8 \pi^2} \sum_t \int_{-\pi/d}^{\pi/d} d k_z \textrm{sign}[m_t(k_z)] \left[1 - \frac{|m_t(k_z)|}{\epsilon_F} \right] \nonumber \\
&\times&\Theta(\epsilon_F - |m_t(k_z)|), 
\eeqa
where $\Theta(x)$ is the Heaviside step function.  
The first term in Eq.~\eqref{eq:15} is the contribution of the completely filled $s = -$ bands. The second term is the contribution 
of incompletely filled $s = +$ bands and, as first pointed out by Haldane,~\cite{Haldane04} can be expressed in terms of an integral of 
the Berry connection field over the Fermi surface. 
Indeed, components of the Berry connection for the $s = +$ bands are given by
\beqa
\label{eq:16}
A^t_x(\bk)&=&-i \langle v^{+ t}_{\bk}| \partial_{k_x} |v^{+ t}_{\bk} \rangle = - \frac{v_F^2 k_y}{2 \epsilon_t(\bk) [\epsilon_t(\bk) + m_t(k_z)]}, \nonumber \\
A^t_y(\bk)&=&-i \langle v^{+ t}_{\bk}| \partial_{k_y} |v^{+ t}_{\bk} \rangle = \frac{v_F^2 k_x}{2 \epsilon_t(\bk) [\epsilon_t(\bk) + m_t(k_z)]}.
\eeqa
Integrating the Berry connection over a 2D section of the Fermi surface, corresponding to a fixed $k_z$, i.e. $v_F^2 (k_x^2 + k_y^2) = \epsilon_F^2 - m_t^2(k_z)$, 
we obtain
\beqa
\label{eq:17}
&&\phi_t(k_z) = \oint d \bk \cdot \bA^{+ t}(\bk) = \pi \left[1 - m_t(k_z)/\epsilon_F\right] \nonumber \\
&=&\pi \left[\textrm{sign}[m_t(k_z)] - m_t(k_z)/\epsilon_F\right] + \pi[1 - \textrm{sign}[m_t(k_z)]], \nonumber \\
\eeqa
where $\phi_t(k_z)$ is the total Berry phase, accumulated along the corresponding section of the Fermi surface. 
Using this, we can rewrite Eq.~\eqref{eq:15} in the following way
\beqa
\label{eq:17.1}
&&\Pi^{inter}(\bq, i\Omega) = \frac{e^2 q}{8 \pi^2} \sum_t  \int_{-\pi/d}^{\pi/d} d k_z \,\, \textrm{sign} [m_t(k_z)]  \nonumber \\
&-&\frac{e^2 q}{8 \pi^2} \sum_t  \int_{-\pi/d}^{\pi/d} d k_z \frac{\phi_t(k_z)}{\pi} \Theta(\epsilon_F - |m_t(k_z)|) \nonumber \\
&+&\frac{e^2 q}{8 \pi^2} \sum_t \int^{\pi/d}_{-\pi/d} d k_z \left[ 1 - \textrm{sign}[m_t(k_z)] \right] \Theta(\epsilon_F - |m_t(k_z)|).  \nonumber \\
\eeqa

The interpretation of Eq.~\eqref{eq:17.1} is, unfortunately, not unambiguous due to the fact that the Berry phase $\phi_t(k_z)$ is only 
defined modulo $2 \pi$. 
Due to this ambiguity, one has some freedom in how to group various terms in Eq.~\eqref{eq:17.1} and assign meaning to them. 
In particular, note that both the first and the last term in Eq.~\eqref{eq:17.1} contributes either $0$ or $2 \pi$ to $\phi_t$. 
Thus we could absorb both terms into $\phi_t$ and obtain
\beq
\label{eq:17.2}
\Pi^{inter}(\bq, i\Omega) = - \frac{e^2 q}{8 \pi^2} \sum_t  \int_{-\pi/d}^{\pi/d} d k_z \frac{\phi_t(k_z)}{\pi} \Theta(\epsilon_F - |m_t(k_z)|), 
\eeq
which means that $\Pi^{inter}(\bq, i \Omega)$ is regarded as entirely a Fermi surface property. This is closely related to Haldane's well-known 
theorem that the anomalous Hall conductivity is a Fermi surface property.~\cite{Haldane04}
Another, equally reasonable, possibility, however, is to absorb only the last term into $\phi_t$. 
In this case we have
\beqa
\label{eq:17.3}
&&\Pi^{inter}(\bq, i\Omega) = \frac{e^2 q}{8 \pi^2} \sum_t  \int_{-\pi/d}^{\pi/d} d k_z\,\,  \textrm{sign} [m_t(k_z)] \nonumber \\
&-&\frac{e^2 q}{8 \pi^2} \sum_t  \int_{-\pi/d}^{\pi/d} d k_z \frac{\phi_t(k_z)}{\pi} \Theta(\epsilon_F - |m_t(k_z)|).
\eeqa
This means that we identify the entire contribution of {\em unfilled} bands with the Fermi surface, while contribution of filled 
bands remains separate. This is the separation employed by us in Ref.~\onlinecite{Burkov13}. 
Finally, perhaps the most natural and most physically motivated interpretation, as will become clear below, results from the following regrouping of terms
\beqa
\label{eq:17.4}
&&\Pi^{inter}(\bq, i\Omega) = \frac{e^2 q}{8 \pi^2} \sum_t  \int_{-\pi/d}^{\pi/d} d k_z \,\, \textrm{sign} [m_t(k_z)] \nonumber \\
&\times&\left[1 - \Theta(\epsilon_F - |m_t(k_z)|)\right] \nonumber \\
&-&\frac{e^2 q}{8 \pi^2} \sum_t  \int_{-\pi/d}^{\pi/d} d k_z \frac{\phi_t(k_z) - \pi}{\pi} \Theta(\epsilon_F - |m_t(k_z)|).\nonumber \\
\eeqa  

To understand the meaning of Eq.~\eqref{eq:17.4}, let us now evaluate the {\em intraband} contribution to $\Pi(\bq, i \Omega)$. 
In this case we have $s = s' = +$ in Eq.~\eqref{eq:9}. It is then clear that, unlike in the case of the interband contribution, evaluated 
above, the value of the intraband contribution depends crucially on the order in which the limits of $\bq  \rightarrow 0$ and $\Omega \rightarrow 0$ 
are taken. If the limit is taken so that $\Omega/ v_F |\bq| \rightarrow \infty$, then one is evaluating the DC limit of a transport quantity (the optical Hall 
conductivity). In this case, the intraband contribution vanishes identically. 
On the other hand, if the limit is taken so that $\Omega/ v_F |\bq| \rightarrow 0$, then one is evaluating an equilibrium thermodynamic property, whose 
physical meaning will become clear below. 
In this case, the intraband contribution is not zero and is given by 
\beqa
\label{eq:18}
\Pi^{intra}(\bq, i\Omega)&=&\frac{i e^2 v_F}{V}\sum_t  \sum_{\bk} \left.\frac{d n_F(x)}{d x} \right|_{x=\epsilon_t(\bk) - \epsilon_F} \nonumber \\
&\times&\langle z^{+ t}_{\bk + \bq} | z^{+ t}_{\bk} \rangle \langle z^{+t}_{\bk} |\bsigma \cdot \hat q| z^{+ t}_{\bk + \bq} \rangle. 
\eeqa 
The derivative of the Fermi distribution function in Eq.~\eqref{eq:18} expresses the important fact that $\Pi^{intra}(\bq, i \Omega)$ is associated 
entirely with the Fermi surface. 
Rotating coordinates so that $\bq = q \hat x$ an expanding to leading order in $q$, as above, we obtain
\beqa
\label{eq:19}
\Pi^{intra}(\bq, i\Omega)&=&- \frac{e^2 v_F^2 q}{2} \sum_t \int \frac{d^3 k}{(2 \pi)^3} \nonumber \\
&\times&\left.\frac{d n_F(x)}{d x} \right|_{x=\epsilon_t(\bk) - \epsilon_F}  \frac{m_t(k_z)}{\epsilon_t^3(\bk)}. 
\eeqa
Evaluating the integral over $k_{x,y}$ as above, we finally obtain
\beqa
\label{eq:20}
&&\Pi^{intra}(\bq, i\Omega) = - \frac{e^2 q}{8 \pi^2} \sum_t \int_{-\pi/d}^{\pi/d} d k_z \, \, \frac{m_t(k_z)}{\epsilon_F} \nonumber \\
&\times&\Theta(\epsilon_F - |m_t(k_z)|) \nonumber \\
&=&\frac{e^2 q}{8 \pi^2} \sum_t  \int_{-\pi/d}^{\pi/d} d k_z \frac{\phi_t(k_z) - \pi}{\pi} \Theta(\epsilon_F - |m_t(k_z)|). \nonumber \\
\eeqa
i.e. the intraband contribution to $\Pi(\bq, i\Omega)$ is equal to the second term in the interband contribution Eq.~\eqref{eq:17.4} in magnitude, 
but opposite in sign. 
Combining the inter- and intraband contributions to $\Pi(\bq, i\Omega)$ we thus obtain 
\beqa
\label{eq:21}
\Pi(\bq, i\Omega)&=&\frac{e^2 q}{8 \pi^2} \sum_t \int_{-\pi/d}^{\pi/d} d k_z \,\, \textrm{sign} [m_t(k_z)]  \nonumber \\
&\times&\left[1 - \Theta(\epsilon_F - |m_t(k_z)|)\right],
\eeqa
i.e. the second term in Eq.~\eqref{eq:17.4} cancels out when 
the low-frequency, long-wavelength limit is taken in such a way that $\Omega/v_F |\bq| \rightarrow 0$. 
On the other hand, when $\Omega/ v_F |\bq| \rightarrow \infty$, the intraband contribution vanishes and 
\beq
\label{eq:22}
\Pi(\bq, i\Omega) = \Pi^{inter}(\bq, i\Omega).
\eeq
This physical difference in the kind of response the system exhibits is precisely the basis of Streda's separation of contributions to the Hall conductivity into $\sigma_{xy}^I$ and $\sigma_{xy}^{II}$.~\cite{Streda} 
Our analysis makes it clear that this in fact the most physically-meaningful separation of contributions to the intrinsic anomalous Hall conductivity. 
\begin{figure}[t]
\subfigure[]{
   \label{fig:1a}
  \includegraphics[width=7cm]{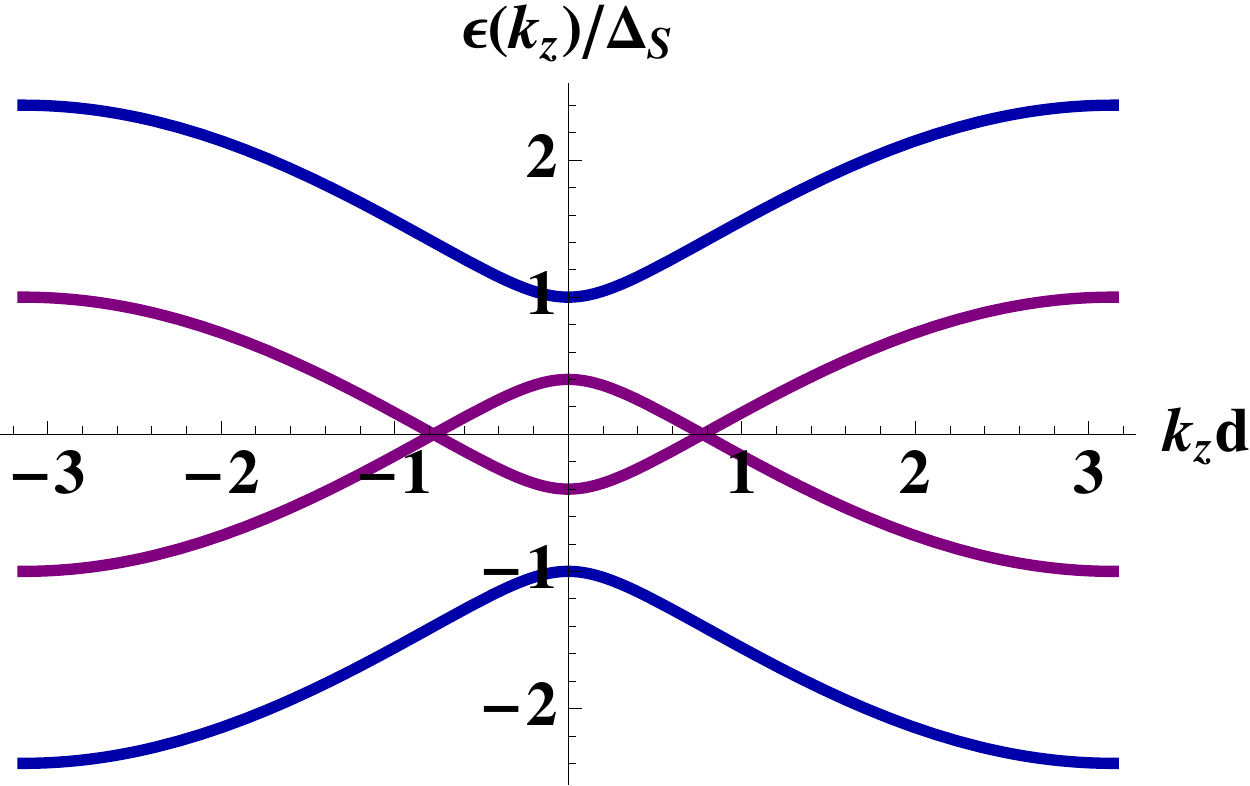}}
\subfigure[]{
  \label{fig:1b}
   \includegraphics[width=7cm]{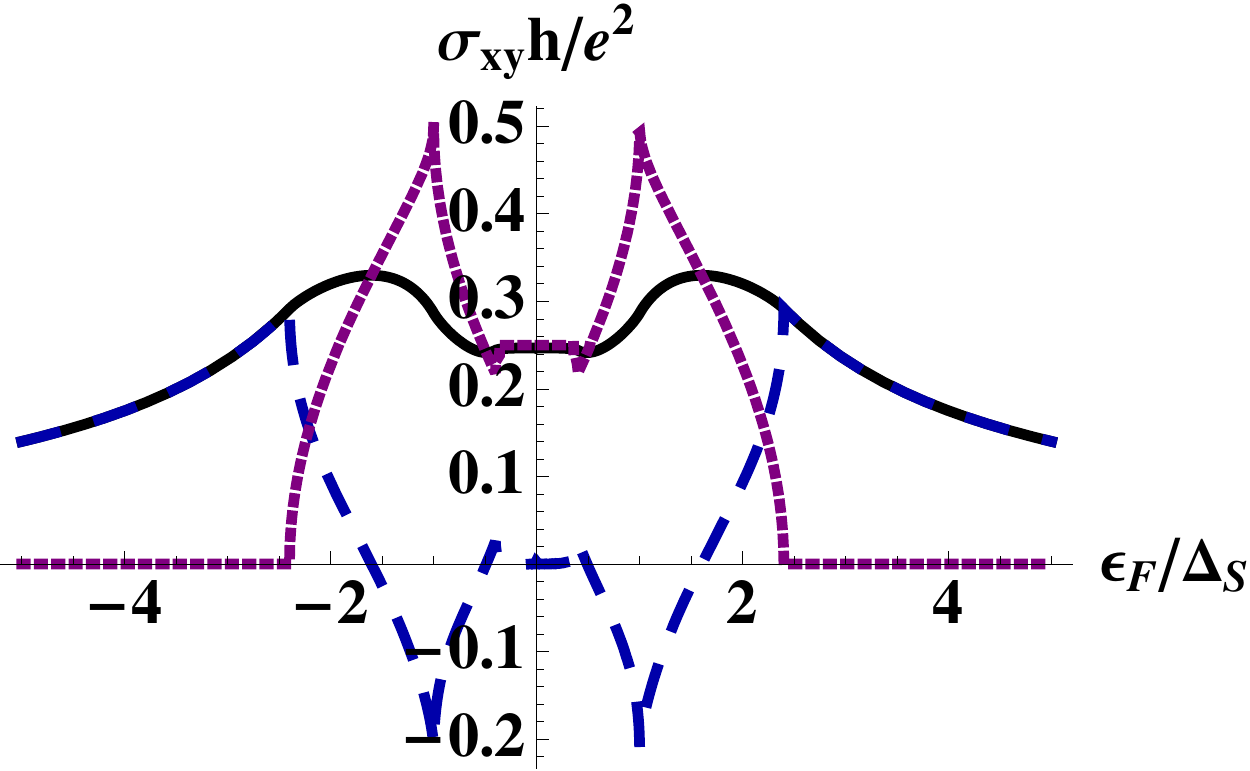}}
  \caption{(Color online). (a) Plot of the band edges along the $z$-direction in momentum space.
  The parameters are such that two Weyl nodes are present.
  (b) Total intrinsic anomalous Hall conductivity (solid line), $\sigma_{xy}^I$ (dashed line), and $\sigma_{xy}^{II}$ (dotted line). Note that the van Hove-like singularities in 
  $\sigma^I_{xy}$ and $\sigma^{II}_{xy}$, associated with band edges, mutually cancel and the total Hall conductivity $\sigma_{xy}$ is a smooth function of the Fermi energy.}
    \label{fig:1}
\end{figure}
\begin{figure}[t]
\subfigure[]{
   \label{fig:2a}
  \includegraphics[width=7cm]{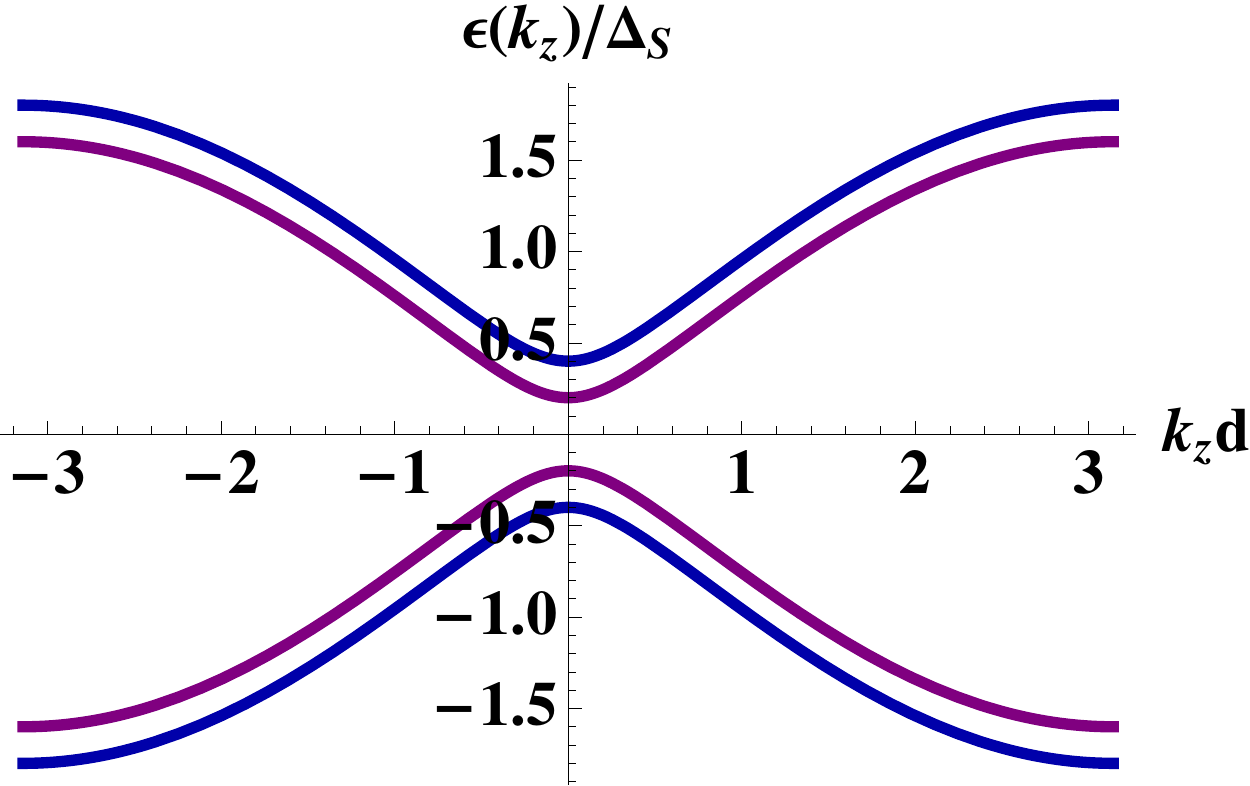}}
\subfigure[]{
  \label{fig:2b}
   \includegraphics[width=7cm]{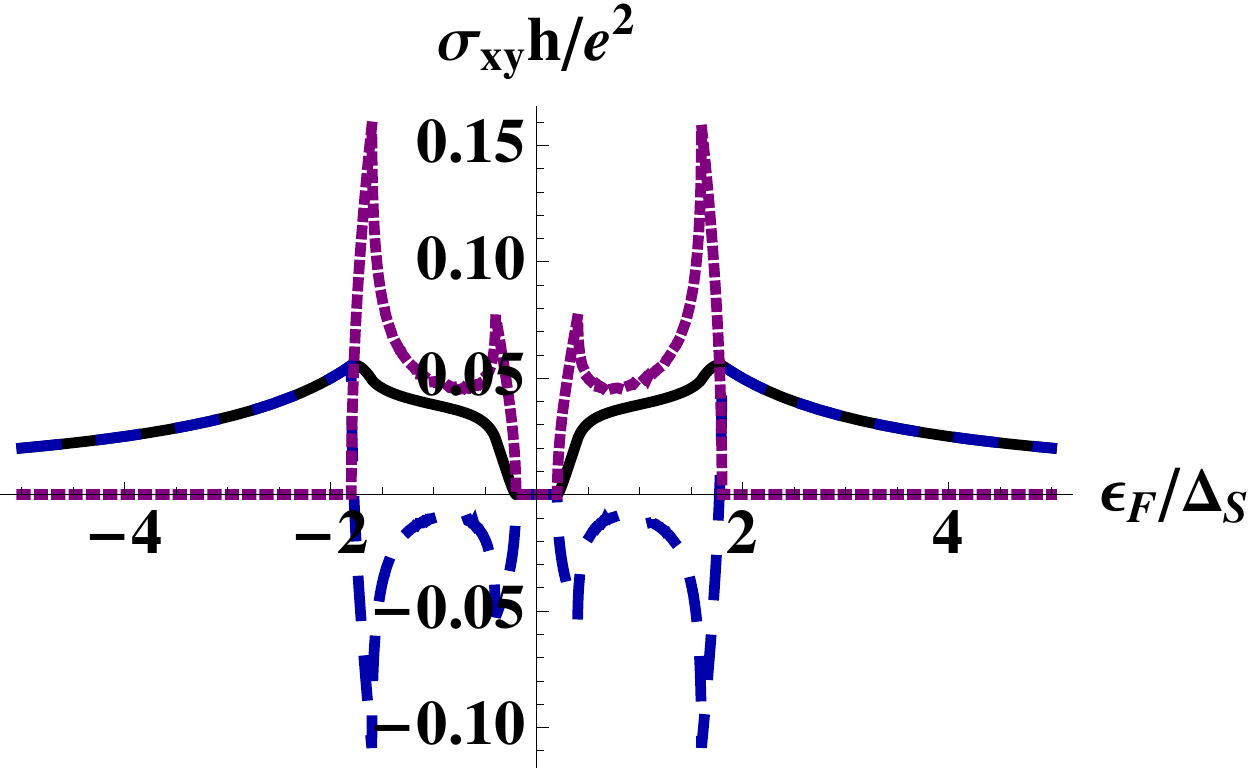}}
  \caption{(Color online). (a) Plot of the band edges along the $z$-direction in momentum space. The spin splitting is not large enough for the 
  Weyl nodes to appear (i.e. $b < |\Delta_S - |\Delta_D||$) and the spectrum has a full gap. 
  (b) Total intrinsic anomalous Hall conductivity (solid line), $\sigma_{xy}^I$ (dashed line), and $\sigma_{xy}^{II}$ (dotted line).}
    \label{fig:2}
\end{figure}

Generalizing the above results to arbitrary sign of $\epsilon_F$ we finally obtain
\beq
\label{eq:23}
S = - i \epsilon^{z 0 \alpha \beta} \sigma_{xy} \int d^3 r d \tau A_0(\br, \tau) \partial_{\alpha} A_{\beta}(\br, \tau), 
\eeq
where, if the low-frequency long-wavelength limit is taken so that $\Omega/ v_F |\bq| \rightarrow \infty$:
\beqa
\label{eq:24}
\sigma_{xy}&=&\frac{e^2}{8 \pi^2} \sum_t \int_{-\pi/d}^{\pi/d} d k_z \,\, \textrm{sign} [m_t(k_z)] \nonumber \\
&\times&\left[\Theta(\epsilon_F + |m_t(k_z)|) - \Theta(\epsilon_F - |m_t(k_z)|)\right] \nonumber \\
&+&\frac{e^2}{8 \pi^2} \sum_t  \int_{-\pi/d}^{\pi/d} d k_z \frac{m_t(k_z)}{|\epsilon_F|} \Theta(|\epsilon_F| - |m_t(k_z)|). \nonumber \\ 
\eeqa
This expression corresponds to the full DC anomalous Hall conductivity. 
On the other hand, when the low-frequency, long-wavelength limit is taken so that $\Omega/ v_F |\bq| \rightarrow 0$, we obtain
\beqa
\label{eq:25}
\sigma_{xy}&=&\sigma_{xy}^{II}=\frac{e^2}{8 \pi^2} \sum_t \int_{-\pi/d}^{\pi/d} d k_z \,\, \textrm{sign} [m_t(k_z)] \nonumber \\
&\times&\left[\Theta(\epsilon_F + |m_t(k_z)|) - \Theta(\epsilon_F - |m_t(k_z)|)\right].
\eeqa
This is precisely the Streda's $\sigma^{II}_{xy}$ contribution to the Hall conductivity, which is a thermodynamic equilibrium quantity, 
equal to
\beq
\label{eq:26}
\sigma^{II}_{xy} = e \left(\frac{\partial N}{\partial B}\right)_{\mu}, 
\eeq
where $N$ is the total electron number. This relation follows immediately from Eq.~\eqref{eq:23} and the order of limits
$\Omega/ v_F |\bq| \rightarrow 0$, which corresponds to thermodynamic equilibrium.
Correspondingly, the $\sigma^{I}_{xy}$ contribution is given by
\beqa
\label{eq:27}
\sigma^I_{xy}&=&\sigma_{xy} - \sigma^{II}_{xy} =  \frac{e^2}{8 \pi^2} \sum_t \int_{-\pi/d}^{\pi/d} d k_z \frac{m_t(k_z)}{|\epsilon_F|} \nonumber \\
&\times&\Theta(|\epsilon_F| - |m_t(k_z)|).
\eeqa
As clear from the above analysis, $\sigma^I_{xy}$ is the contribution to $\sigma_{xy}$ that can be associated with 
states on the Fermi surface. This contribution is nonuniversal, i.e. it depends on details of the electronic structure and 
is a continuous function of the Fermi energy. 
$\sigma^{II}_{xy}$, on the other hand, is the contribution of all states below the Fermi energy and is a thermodynamic 
equilibrium property of the ferromagnet. It attains a universal value,  which depends only on the distance between the 
Weyl nodes, when the Fermi energy coincides with the nodes, i.e. when $\epsilon_F = 0$
\beq
\label{eq:28}
\sigma_{xy}^{II} = \frac{e^2 {\cal K}}{4 \pi^2},
\eeq
where 
\beq
\label{eq:30}
{\cal K} = \frac{2}{d} \arccos\left(\frac{\Delta_S^2 + \Delta_D^2 - b^2}{2 \Delta_S |\Delta_D|} \right), 
\eeq
is the distance between the Weyl nodes. 
When $b > |\Delta_S + |\Delta_D||$, the Weyl nodes annihilate at the edges of the Brillouin zone and a gap opens up. 
In this case ${\cal K} = 2 \pi/d$, i.e. a reciprocal lattice vector and $\sigma^{II}_{xy}$ is quantized as long as the Fermi level is in the gap.~\cite{Halperin92} 
Both contributions, along with the total anomalous Hall conductivity $\sigma_{xy}$ are plotted as a function of the Fermi energy in Figs.~\ref{fig:1},~\ref{fig:2}
in two different cases: when Weyl nodes are present and when they are not. 
The former occurs when $|\Delta_S - |\Delta_D|| < b < |\Delta_S + |\Delta_D||$. 
As can be seen from Fig.~\ref{fig:1}, Weyl nodes provide the dominant contribution to $\sigma^{II}_{xy}$ and to the total Hall conductivity $\sigma_{xy}$, if the 
Fermi level is not too far from the nodes. 

It is also worthwhile to note the following interesting property, which is evident from Fig.~\ref{fig:1}. 
Both the total anomalous Hall conductivity $\sigma_{xy}$ and the two distinct contributions to it, $\sigma_{xy}^{I,II}$ appear to exhibit a 
quasi-plateau behavior when $\epsilon_F$ is not too far from the Weyl nodes. To understand the origin of this behavior, consider the derivative 
of $\sigma^{II}_{xy}$ with respect to the Fermi energy
\beq
\label{eq:31}
\frac{\partial \sigma^{II}_{xy}}{\partial \epsilon_F} = - \frac{e^2}{8 \pi^2} \sum_t \int_{-\pi/d}^{\pi/d} d k_z \textrm{sign}[m_t(k_z)] \delta(\epsilon_F - |m_t(k_z)). 
\eeq
Unlike Eq.~\eqref{eq:25}, this is straightforward to evaluate analytically. We obtain
\beqa
\label{eq:32}
\frac{\partial \sigma_{xy}^{II}}{\partial \epsilon_F}&=&- \frac{e^2}{8 \pi^2} \int_{-\pi/d}^{\pi/d} d k_z \nonumber \\
&\times&\left[\delta(\Delta(k_z) - b + \epsilon_F) - \delta(\Delta(k_z) - b -\epsilon_F) \right] \nonumber \\
&=&\frac{e^2}{4 \pi^2}\left(1/\tilde v_{F+} - 1/\tilde v_{F-} \right), 
\eeqa
where we have assumed that $\epsilon_F$ is sufficiently close to zero, so that only the $t = -$ bands contribute to the integral, and 
\beqa
\label{eq:33}
&&\tilde v_{F\pm} = \frac{d}{2 (b \pm \epsilon_F)} \nonumber \\
&\times&\sqrt{[(b \pm \epsilon_F)^2 - (\Delta_S - |\Delta_D|)^2] [(\Delta_S + |\Delta_D|)^2 - (b \pm \epsilon_F)^2]}, \nonumber \\
\eeqa
are the two Fermi velocities, corresponding to two pairs of solutions of the equation $|b - \Delta(k_z)| = \epsilon_F$. 
It is then clear that as long as $\Delta_S - |\Delta_D| \ll b \pm \epsilon_F \ll \Delta_S + \Delta_D$, both Fermi velocities are independent 
of the Femi energy and thus $\partial \sigma_{xy}^{II}/\partial \epsilon_F$ vanishes. This simply means that as long as the band dispersion 
near the Weyl nodes may be regarded as linear, the Fermi velocity is independent of the Fermi energy. 
By a nearly identical calculation it is easy to show that $\partial \sigma^{I}_{xy} / \partial \epsilon_F$ also vanishes when $\epsilon_F$ is sufficiently close to zero. 
This is the origin of the quasi-plateau behavior in Fig.~\ref{fig:1}.
This result implies that the intrinsic anomalous Hall conductivity is equal to its thermodynamic equilibrium part, $\sigma^{II}_{xy}$, not just when the Fermi 
energy coincides with the Weyl nodes, but even away from them as long as the band dispersion may be assumed to be linear. 
\section{Discussion and conclusions}
\label{sec:3}
The main message of this paper is that it is possible to separate out two distinct contributions to the intrinsic anomalous Hall conductivity:
a contribution that can be associated with states on the Fermi surface, $\sigma^I_{xy}$, and a contribution that can not be assigned to states on the Fermi surface and is instead 
associated with all filled states, $\sigma_{xy}^{II}$. 
This separation coincides with the one proposed long time ago by 
Streda~\cite{Streda} in the context of the quantum Hall effect. The physical basis for this is that while $\sigma^{II}_{xy}$ is a thermodynamic 
equilibrium property of the material, the Fermi surface contribution $\sigma^I_{xy}$ is a purely transport property, which disappears in equilibrium. 
This physical distinction between the two contributions may, in principle, allow one to separate them experimentally.  

$\sigma^I_{xy}$ and $\sigma^{II}_{xy}$ will also be affected very differently by disorder. The role of impurity scattering in the AHE has long been a 
highly controversial issue.~\cite{Nagaosa10} 
What complicates matters especially is the existence of the so-called side-jump contribution to the anomalous Hall conductivity, which, while 
arising from impurity scattering, is, paradoxically, independent of the impurity concentration. This contribution is thus always of the same order 
as the intrinsic contribution and may even cancel it exactly in some models.~\cite{MacDonald07,Niu11} 
We can expect, however, that only the Fermi surface part of the intrinsic anomalous Hall conductivity $\sigma^I_{xy}$ will be significantly affected 
by the impurity scattering. The thermodynamic equilibrium contribution $\sigma^{II}_{xy}$ should be, at least approximately, independent of disorder,
making its role especially important in real materials. We leave detailed investigation of these issues to future work. 

Finally, it would also be interesting to extend the above considerations to more realistic models of metallic ferromagnets, such as the model 
discussed by us before in Ref.~\onlinecite{Burkov13}. 

\begin{acknowledgments}
Financial support was provided by NSERC of Canada and a University of Waterloo start up grant. 
\end{acknowledgments}


\begin{thebibliography}{99}
\bibitem{TI} C.L. Kane and E.J. Mele, Phys. Rev. Lett. {\bf 95}, 146802 (2005); 
B.A. Bernevig, T.L. Hughes, and S.-C. Zhang, Science {\bf 314}, 1757 (2006); 
J.E. Moore and L. Balents, Phys. Rev. B {\bf 75}, 121306 (2007); 
R. Roy, Phys. Rev. B {\bf 79}, 195322 (2009); 
L. Fu, C.L. Kane, and E.J. Mele, Phys. Rev. Lett. {\bf 98}, 106803 (2007);
 M. K\"onig, S. Wiedmann, C. Br\"une, A. Roth, H. Buhmann, L. Molenkamp, 
X.-L. Qi, and S.-C. Zhang, Science {\bf 318}, 766 (2007); Y. Xia, D. Qian, D. Hsieh, L. Wray, A. Pal, H. Lin, A. Bansil,  D. Grauer, Y.S. Hor, R.J. Cava, 
and M.Z. Hasan, Nature Phys. {\bf 5}, 398 (2009); 
M.Z. Hasan and C.L. Kane, Rev. Mod. Phys. {\bf 82}, 3045 (2010); X.-L. Qi and S.-C. Zhang, Rev. Mod. Phys. {\bf 83}, 1057 (2011).  
\bibitem{Volovik} G.E. Volovik, {\em The Universe in a Helium Droplet} (Clarendon Press, Oxford, 2003); Lect. Notes Phys. {\bf 718}, 31 (2007).
\bibitem{Wan11} X. Wan, A.M. Turner, A. Vishwanath, and S.Y. Savrasov, Phys. Rev. B {\bf 83}, 205101 (2011). 
\bibitem{Ran11} K.-Y. Yang, Y.-M. Lu, and Y. Ran, Phys. Rev. B {\bf 84}, 075129 (2011).
\bibitem{Burkov11} A.A. Burkov and L. Balents, Phys. Rev. Lett. {\bf 107}, 127205 (2011).
\bibitem{Xu11} G. Xu, H.-M. Weng, Z.-J. Wang, X. Dai, and Z. Fang, Phys. Rev. Lett. {\bf 107}, 186806 (2011).
\bibitem{Aji11} V. Aji, Phys. Rev. B {\bf 85}, 241101 (2012). 
\bibitem{Burkov12}  A.A. Zyuzin and A.A. Burkov, Phys. Rev. B {\bf 86}, 115133 (2012).
\bibitem{Grushin12} A.G. Grushin, Phys. Rev. D {\bf 86}, 045001 (2012). 
\bibitem{Goswami13} P. Goswami and S. Tewari, Phys. Rev. B {\bf 88}, 245107 (2013). 
\bibitem{Pesin13}  S.A. Parameswaran, T. Grover, D.A. Abanin, D.A. Pesin, and A. Vishwanath, arXiv:1306.1234 (unpublshed). 
\bibitem{Cava13} S. Borisenko, Q. Gibson, D. Evtushinsky, V. Zabolotnyy, B. B\"uchner, and R.J. Cava, arXiv:1309.7978 (unpublished). 
\bibitem{Shen13} Z.K. Liu, B. Zhou, Z.J. Wang, H.M. Weng, D. Prabhakaran, S.-K. Mo, Y. Zhang, Z.X. Shen, Z. Fang, X. Dai, Z. Hussain, and Y.L. Chen, 
arXiv:1310.0391 (unpublished). 
\bibitem{Hasan13} M. Neupane, S.-Y. Xu, R. Sankar, N. Alidoust, G. Bian, C. Liu, I. Belopolski, T.-R. Chang, H.-T. Jeng, H. Lin, A. Bansil, F. Chou, 
and M.Z. Hasan, arXiv:1309.7892 (unpublished). 
\bibitem{Burkov13} Y. Chen, D.L. Bergman, and A.A. Burkov, Phys. Rev. B {\bf 88}, 125110 (2013). 
\bibitem{Karplus54} R. Karplus and J.M. Luttinger, Phys. Rev. {\bf 95}, 1154 (1954); 
J. M. Luttinger, Phys. Rev. {\bf 112}, 739 (1958).
\bibitem{Niu99} G. Sundaram and Q. Niu, Phys. Rev. B {\bf 59}, 14915 (1999).  
\bibitem{MacDonald02} T. Jungwirth, Q. Niu, and A.H. MacDonald, Phys. Rev. Lett. {\bf 88}, 207208 (2002). 
\bibitem{Nagaosa02} M. Onoda and N. Nagaosa, J. Phys. Soc. Japan {\bf 71}, 19 (2002). 
\bibitem{Fang03} Z. Fang, N. Nagaosa, K.S. Takahashi, A. Asamitsu, R. Mathieu, T. Ogasawara, H. Yamada, M. Kawasaki, Y. Tokura, and K. Terakura, Science {\bf 302}, 92 (2003).
\bibitem{MacDonald04} Y. Yao, L. Kleinman, A.H. MacDonald, J. Sinova, T. Jungwirth, D.S. Wang, E. Wang, and Q. Niu, Phys. Rev. Lett. {\bf 92}, 037204 (2004). 
\bibitem{Haldane04} F.D.M. Haldane, Phys. Rev. Lett. {\bf 93}, 206602 (2004).
\bibitem{Ong04} W.-L. Lee, S. Watauchi, R.J. Cava, and N.P. Ong, Science {\bf 303}, 1647 (2004). 
\bibitem{Nagaosa10} N. Nagaosa, J. Sinova, S. Onoda, A.H. MacDonald, and N.P. Ong, Rev. Mod. Phys. {\bf 82}, 1539 (2010). 
\bibitem{Streda} P. Streda, J. Phys. C {\bf 15}, L717 (1982). 
\bibitem{Halperin92} M. Kohmoto, B.I. Halperin, and Y.S. Wu, Phys. Rev. B {\bf 45}, 13488 (1992). 
\bibitem{MacDonald07} N.A. Sinitsyn, A.H. MacDonald, T. Jungwirth, V.K. Dugaev, and J. Sinova, Phys. Rev. B {\bf 75}, 045315 (2007). 
\bibitem{Niu11} S.A. Yang, Z. Qiao, Y. Yao, J. Shi, and Q. Niu, Europhys. Lett. {\bf 95}, 67001 (2011). 
\end{thebibliography}
\end{document}